\documentclass{article} 
\usepackage{iclr2026_conference,times}

\usepackage{amsmath,amsfonts,bm}

\def\eqref#1{equation~\ref{#1}}

\def\1{\bm{1}}

\DeclareMathAlphabet{\mathsfit}{\encodingdefault}{\sfdefault}{m}{sl}
\SetMathAlphabet{\mathsfit}{bold}{\encodingdefault}{\sfdefault}{bx}{n}

\usepackage{hyperref}
\usepackage{url}
\usepackage{fancyhdr}

\title{Decoupling Identity from Utility: Privacy-by-Design Frameworks for Financial Ecosystems}

\author{Ifayoyinsola Ibikunle*, Tyler Farnan, Senthil Kumar \& Mayana Pereira  \\
Capital One\\
}

\fancypagestyle{firstpage}{
    \fancyhf{}
    \lhead{Accepted as a Financial AI workshop paper at ICLR 2026}
    \lfoot{Correspondence to Ifayoyinsola Ibikunle (ifayoyinsola.ibikunle@capitalone.com)} 

}

\iclrfinalcopy 
\begin{document}
\maketitle
\thispagestyle{firstpage}
\begin{abstract}
Financial institutions face tension between maximizing data utility and mitigating the re-identification risks inherent in traditional anonymization methods. This paper positions Differentially Private (DP) synthetic data as a robust ``Privacy-by-Design'' framework for building \textbf{Responsible Agentic Systems} in Finance. We examine two distinct generative paradigms: (1) Direct Tabular Synthesis, which reconstructs high-fidelity joint distributions from raw data, and (2) DP-Seeded Agent-Based Modeling (ABM), which uses DP-protected aggregates to parameterize complex, stateful simulations. We argue that while tabular synthesis suffices for static analytics, DP-Seeded ABM is essential for the era of autonomous finance. It provides a ``safe gym'' for training agents, enabling fairness auditing through counterfactual parameterization and robustness testing against simulated ``black swan'' events, all while adhering to rigorous formal privacy guarantees.
\end{abstract}

\section{Introduction}

Financial data is the cornerstone of the global economic system, holding significant value for institutional operations, research, and regulatory oversight. However, leveraging this data creates a fundamental challenge: increased data utility is subject to increased risk to individual privacy. Traditional anonymization (e.g., k-anonymity) offers limited defense against re-identification attacks \citep{e380c6c9215446d1b4590baae9708d11, 10.1197/jamia.M2716, ohm}. True privacy requires a decoupling of individual identities from data utility.  

To resolve this, the industry is shifting towards Privacy-Enhancing Technologies (PETs), specifically Differentially Private (DP) synthetic data \citep{WuNan2020, chen2021financialvisionbaseddifferential, FeiWang2024}. Differential Privacy is a robust mathematical framework that provides a formal guarantee of privacy by injecting calibrated noise into statistical queries from a dataset \citep{10.1007/11681878_14, 10.1561/0400000042}. This ensures that the addition or removal of any single record has no perceptible effect on the final output. 

\subsection{Related Works}

\textbf{Foundations of Private Data Synthesis:} Differentially Private Synthetic Data (DPSD) for sensitive domains is commonly generated generated through three primary paradigms: marginal-based algorithms (e.g., AIM \citep{mckenna2024aimadaptiveiterativemechanism} and MST \citep{mckenna2021winningnistcontestscalable} ), probabilistic graphical models (e.g., PrivBayes \citep{zhang2017privbayes}), and neural generative architectures(e.g., PATE-GAN \citep{jordon2019pategan}). These approaches provide formal privacy guarantees and strong marginal fidelity, but often struggle to preserve long-horizon state, domain constraints, and causal mechanisms without substantial architectural or post-hoc intervention. 

In these contexts, the rigor of \textit{privacy} is governed by the parameter $\epsilon$. Although a smaller $\epsilon$ improves data privacy, it presents a utility-privacy trade-off where excessive noise overwhelms the underlying signal in the data, degrading higher-order dependencies and rare-event behavior \citep{ghosh2009universally, lu2014generating, pmlr-v108-geng20a}. Empirical work further highlights that this tension disproportionately impacts minority subpopulations, leading to uneven downstream performance \citep{stadler2022groundhog, ganev2022robinhood}. Such limitations are especially problematic for financial event sequences, where mechanistic state updates and causal decision dynamics are critical for realistic synthesis. 

\textbf{From Static Synthesis to State-Aware Simulation:} While traditional synthesis focuses on data snapshots, rules-based and agent-based simulators generate data through explicit mechanisms and state transitions. This approach enables constraint-consistent realism and controllable counterfactuals by construction. These types of simulators are widely used in finance due to data sensitivity and strong logical requirements, but are typically calibrated using non-private aggregates \citep{lopezrojas2014banksim, lopezrojas2016paysim}. A notable exception is MoMTSimDP, which calibrates a mobile money simulator using DP summary statistics \citep{azamuke2025momtsimdp}. 

\textbf{Benchmarking Environments for Financial Agents:}
Building upon these simulation capabilities, the training of autonomous agents has been accelerated by open-source environments like \textbf{FinRL} \citep{liu2021finrl} and \textbf{ABIDES} \citep{Amrouni_2021, dwarakanath2024abides}. These platforms provide standardized ``gyms'' for training Deep Reinforcement Learning (DRL) agents in complex market conditions. Frameworks like \textbf{FinMem} \citep{yu2024finmem} have advanced the cognitive capabilities of financial agents through profiling and memory retrieval mechanisms to allow agents to reason through past experiences. 

While these frameworks have successfully enabled the development of autonomous trading strategies, it predominantly relies on public market data (e.g., stock prices, news sentiment). To bridge this domain, PriMORL \citep{rio2024primorl} introduces a model-based Reinforcement Learning (RL) with formal DP guarantees. By constructing a privacy-preserving surrogate of the underlying environment, PriMORL enables the training of agents on sensitive offline continuous control tasks with deep function approximations. Together, these advancements provide the infrastructure and mathematical foundation for autonomous financial agents.

\textbf{Bridging the Privacy Gap for Autonomous Finance:}
Directly adapting these powerful architectures to retail banking or fraud detection poses severe privacy risks. Training agents on raw transaction logs risks memorization of user identities, a vulnerability not addressed by standard trading gyms. Current anonymization methods often rely on ad-hoc masking, which fails to provide formal guarantees against linkage attacks \citep{azamuke2025momtsimdp}. This paper argues for merging the rigorous ``Privacy-by-Design'' calibration of MoMTSimDP \citep{azamuke2025momtsimdp} with the agentic infrastructure of frameworks like FinRL and ABIDES to create \textit{Differentially Private Gyms} for sensitive financial use cases.


\textbf{Taxonomy of Privacy in the Data Sharing Lifecycle:} 
To evaluate PET efficacy, it is  important to distinguish between the protections applied during computation \textit{(input privacy)} and those applied to the resulting artifact \textit{(output privacy)}. Input privacy \citep{Kantarcioglu04, Sangeetha2019}, as explored in \cite{chatzigiannis2023privacyenhancingtechnologiesfinancialdata}, utilizes protocols like homomorphic encryption \citep{armknecht2015guide} and secure multi-party computation \citep{4568388, 10.1145/237814.238015, Kantarcioglu04} to secure raw data during processing, enabling collaborative tasks like anti-money laundering detection without data disclosure. Output privacy \citep{Sangeetha2019, bu2006preservation}, on the other hand, ensures that the final synthetic artifact (e.g., statistical summary, synthetic dataset, or machine learning model) does not leak information about specific individuals in the dataset. This paper deviates from the secure computation focus of \cite{chatzigiannis2023privacyenhancingtechnologiesfinancialdata} and centers on output privacy. By prioritizing Differential Privacy (DP) at the output stage, we ensure that the resulting synthetic ``gyms'' are inherently safe for external distribution, regardless of the underlying computational path.

\section{Generative Paradigms for Privacy-Preserving Financial Data}
We compare two distinct paradigms for overcoming the structural hurdles inherent in financial data; where the high-dimensionality of relational records and the complex temporal dependencies of transactional histories collide. The key distinctions between these modalities are evidenced in Table~\ref{Paradigm-comparision}, which contrasts the static nature of traditional synthesis with the mechanistic, rule-based evolution of a DP-Seeded ``gym'' environment across a number of critical dimensions. 

\subsection{Direct Tabular Synthesis (Static)}
Direct synthesis employs DP algorithms to ingest raw data and reconstruct joint distributions into a fixed table. This methodology establishes a gold standard for static use cases, such as QA testing or marketing analytics, where fidelity to a specific historical window is paramount. However, it fails to capture the \textit{counterfactual} risks required for robust agent training.

\subsection{DP-Seeded Agent-Based Modeling (Dynamic)}
This paper identifies DP-Seeded ABM as a robust evolution for agentic workflows. Under this framework, DP is not applied to the final output, but to the \textit{input parameters} of a simulation engine. For example, MoMTSimDP computes DP-protected statistics on client profiles (e.g., transaction frequency, maximum balance) and aggregated metrics (e.g., hourly transaction counts) \citep{azamuke2025momtsimdp}. These noisy statistics then ``seed'' the stochastic behavior of simulated agents.

This approach decouples individual identities from the simulation logic. Even if an adversary interacts with the simulation indefinitely, the underlying privacy guarantee is anchored by the budget used to generate the seed statistics.

\begin{table}[htpb]
    \caption{The paradigm shift: From data to environments. The distinctions shown between these generative data approaches demonstrates that moving towards mechanistic, rule-based logic transforms synthetic data from a static resource into a dynamic, high-fidelity training ``gym''. By embedding privacy in the environmental parameters rather than the raw output, this approach allows for the generation of complex, stateful simulations essential for autonomous agent development within a risk-mitigated framework.} 
    \label{Paradigm-comparision}
    \begin{center}
        \begin{tabular}{p{7em}|p{14em}|p{12em}}
        \multicolumn{1}{c}{\bf Feature}  &\multicolumn{1}{c}{\bf Direct Tabular Synthesis} &\multicolumn{1}{c}{\bf DP-Seeded ABM} \\
        \hline 
        Logic Source & Statistical Correlations & Mechanistic Rules \\
        \hline 
        Temporal State & Static / Fixed Window & Stateful Evolution \\
        \hline 
        Privacy Locus & Output Data & Environment Parameters \\
        \hline 
        Primary Utility & Analytics \& QA Testing & Agent Training \& Evaluation \\
        \hline
        \end{tabular}
    \end{center}
\end{table}

\section{Enabling Responsible Autonomy in Financial Systems}

The transition from static tabular synthesis to DP-Seeded ABM is a prerequisite for the safe deployment of agentic systems. While frameworks like FinRL \citep{liu2021finrl} serve algorithmic trading, no equivalent exists for sensitive ``retail'' tasks due to privacy risks. We advocate for the \textit{DP-Seeded Gym} as the standardized infrastructure needed to bridge this gap, enabling three critical workflows:

\subsection{Democratizing ``Retail'' RL \& Fairness Calibration}
Training agents on historical data risks automating embedded biases~\citep{azamuke2025momtsimdp}. A DP-Seeded ABM allows researchers to intervene at the calibration level—adjusting ``seed'' statistics to oversample underrepresented groups or normalize transaction frequencies. This creates a ``fair world'' baseline where agents can learn equitable policies before deployment, creating open-source ``Retail Gyms'' compatible with advanced architectures like FinMem \citep{yu2024finmem} or ABIDES \citep{Amrouni_2021} without compromising customer privacy.

\subsection{Benchmarking Robustness in ``Black Swan'' Scenarios}
Responsible agents must remain stable during market shocks, yet static datasets lack sufficient ``tail events.'' A DP-Seeded ABM acts as a ``counterfactual laboratory''~\citep{azamuke2025momtsimdp}. By perturbing privacy-protected parameters—such as increasing transaction volatility or altering interaction probabilities~\citep{azamuke2025momtsimdp}—regulators can generate synthetic stress scenarios (e.g., bank runs). This enables standardized test suites where agents are audited against thousands of unseen, extreme economic conditions.

\subsection{Multi-Agent Adversarial Training}
Modern fraud detection requires cooperative multi-agent systems that adapt to evolving threats \citep{deloitte2025agentic}. A DP-calibrated simulator supports this by allowing the injection of parametrized ``fraudster'' agents~\citep{azamuke2025momtsimdp}. Because honest agent profiles are derived from DP statistics~\citep{azamuke2025momtsimdp}, the system simulates realistic customer friction in response to aggressive fraud checks, enabling the training of robust defensive agents without exposing real user behavior.

\section{Technical Challenges: The Privacy-Fidelity Frontier}

Constructing ``Privacy-Safe Gyms'' introduces distinct technical hurdles that differentiate them from standard synthetic data tasks:

\textbf{Simulation Drift \& Error Propagation:} Unlike tabular synthesis, where DP noise is added once, ABMs use noisy parameters to drive recursive time-steps. MoMTSimDP, for instance, executes for 720 sequential steps where agent states (e.g., account balances) persist and influence future interactions~\citep{azamuke2025momtsimdp}. In such recursive systems, small privacy-induced errors in the initial calibration can compound over the simulation horizon, potentially causing the synthetic economy to drift into unrealistic macroeconomic states (e.g., hyperinflation). Future work must focus on \textit{noise-aware calibration}, where agent logic explicitly accounts for the variance inherent in the DP statistics used to seed the model.

\textbf{The Sim-to-Real Gap in RL:} Agents trained in simulations often fail in the real world due to overfitting to the simulator's artifacts. Currently, DP-Seeded simulators like MoMTSimDP validate utility using Sum of Squared Errors (SSE) on marginal distributions~\citep{azamuke2025momtsimdp}.  While sufficient for static analysis, this metric fails to capture \textit{interaction fidelity}. Future work requires new benchmarks that measure whether the causal dynamics (e.g., an agent's sensitivity to transaction fees) in the private simulation match those in the sensitive, real-world environment.

\section{Regulatory Alignment/Future Outlook}
By moving from static data synthesis to DP-Seeded environment generation, we can responsibly unlock the potential of Agentic AI for sensitive use cases in consumer finance. Differential Privacy ensures testing and analytics remain resilient against the ever-changing regulatory and ethical landscape by achieving a ``Privacy-by-Design'' approach. Looking forward, the output-privacy paradigms presented here empower analysts and external researchers to derive high-utility insights without traditional data-clearing bottlenecks. Ultimately, these privacy-driven datasets enable seamless cross-organization collaboration, ensuring that data-driven decision-making remains agile and compliant. 

\newpage

\bibliography{iclr2026_conference}
\bibliographystyle{iclr2026_conference}

\end{document}